\documentclass{article}
\usepackage{spconf,amsmath,graphicx}
\usepackage{enumitem}
\usepackage{hyperref}
\usepackage{graphicx} 
\usepackage{array}  
\usepackage{booktabs}
\setlist{nosep, leftmargin=14pt}
\usepackage{mwe} 

\title{Efficient Lung Ultrasound Severity Scoring Using Dedicated Feature Extractor}

\name{\begin{tabular}{c}
Jiaqi Guo$^{1\dagger}$ \thanks{© 20XX IEEE. Personal use of this material is permitted. Permission from IEEE must be obtained for all other uses, in any current or future media, including reprinting/republishing this material for advertising or promotional purposes, creating new collective works, for resale or redistribution to servers or lists, or reuse of any copyrighted component of this work in other works.}, Yunan Wu$^{1}$, Evangelos Kaimakamis$^{2}$, Georgios Petmezas$^{3}$, Vasileios E. Papageorgiou$^{3}$, \\Nicos Maglaveras$^{3}$, Aggelos K. Katsaggelos$^{1\mathcal{y}}$~\thanks{$^{\dagger}$ Corresponding author.}
\end{tabular}}

\address{$^{1}$ Northwestern University, Illinois, USA\\
$^{2}$ "G. Papanikolaou" General Hospital, Thessaloniki, GREECE\\
$^{3}$ Aristotle University, Thessaloniki, GREECE}

\begin{document}
\maketitle
\begin{abstract}
With the advent of the COVID-19 pandemic, ultrasound imaging has emerged as a promising technique for COVID-19 detection, due to its non-invasive nature, affordability, and portability. In response, researchers have focused on developing AI-based scoring systems to provide real-time diagnostic support. However, the limited size and lack of proper annotation in publicly available ultrasound datasets pose significant challenges for training a robust AI model. This paper proposes MeDiVLAD, a novel pipeline to address the above issue for multi-level lung-ultrasound (LUS) severity scoring. In particular, we leverage self-knowledge distillation to pretrain a vision transformer (ViT) \textit{without label} and aggregate frame-level features via \textit{dual-level VLAD aggregation}. We show that with minimal finetuning, MeDiVLAD outperforms conventional fully-supervised methods in both frame- and video-level scoring, while offering classification reasoning with exceptional quality. This superior performance enables key applications such as the automatic identification of critical lung pathology areas and provides a robust solution for broader medical video classification tasks.\footnote{Code: \href{https://github.com/GuoJiaqi-1020/MeDiVLAD}{\path{https://github.com/GuoJiaqi-1020/MeDiVLAD}}} 
\end{abstract}
\begin{keywords}
Deep Neural Network, DINO, VLAD, Medical Video Classification, Lung Ultrasound Score
\end{keywords}

\section{Introduction}
\label{sec:intro}
The LUS score is a crucial tool for assessing lung disease severity~\cite{volpicelli2021lung}, particularly helpful for non-expert practitioners in evaluating patients with pulmonary abnormalities in unsupervised settings. AI-based LUS scoring typically involves extracting frame-level features with a pretrained deep neural network (DNN), then aggregating them into video-level embeddings. This method faces two challenges: \textbf{Network Pretraining}: Limited public LUS datasets, varying acquisition systems, image quality, and lack of annotations hinder the effectiveness of traditional supervised learning. \textbf{Frame-level Aggregation}: Aggregating frame-level features risks information loss, making it crucial to retain key details for optimal LUS performance. To address the first challenge, several AI-based scoring methods leverage contrastive learning~\cite{roy2020deep, xue2021modality, gare2022weakly}, which learns discriminative features by promoting intra-class similarity. However, most of these approaches rely on costly expert annotations for training. Interestingly, we identify a similar solution in self-supervised learning (SSL) methods, which can be trained without labels and have demonstrated potential in image processing~\cite{caron2020unsupervised, chen2020simple, grill2020koray, caron2021emerging}. Among these methods, DINO~\cite{caron2021emerging} stands out by leveraging a Vision Transformer (ViT) to learn representations through self-knowledge distillation, enabling the training of self-attention mechanisms with limited data. This makes DINO an ideal candidate for pretraining feature extractors in frame-level tasks.

For the second challenge, a straightforward method is to apply the maximum frame-level probability across the video~\cite{gare2022weakly}, classifying based on the most prominent frame in the video. Other frame-level feature aggregations in video classification are commonly addressed using two paradigms. The first employs recurrent neural networks (RNNs)~\cite{petmezas2024recent} to model the temporal dynamics of video sequences, deriving an overall representation from the frame-level features. The second paradigm, broadly categorized as the Bag-of-Visual-Words (BoVW) based aggregation, constructs local frame descriptors, assigns them to pre-defined clusters, and aggregates their residuals into a global representation. A common example of this approach is VLAD~\cite{jegou2010aggregating} and its variants~\cite{arandjelovic2016netvlad,lin2018nextvlad}. Other methods, such as I3D~\cite{carreira2017quo}, which relies on 3D convolutions, are computationally expensive. In comparison, we believe BoVW-based aggregation will be more suitable for our task than RNN-based methods, as LUS video labels often depend on several typical frames within a short time frame, making temporal dependencies less significant.

In this paper, we proposed a semi-self-supervised learning pipeline, \textbf{MeDiVLAD}, for accurate LUS severity scoring. Our method utilizes self-knowledge distillation to pretrain a ViT backbone without labels and finetune it with minimal supervision. To capture both temporal and spatial patterns, we introduce a VLAD-based dual-level assignment matrix for aggregating frame-level features. Remarkably, even without supervised finetuning, the pretrained ViT surpasses a fully supervised ResNet50 on the frame-level LUS task, with its attention map demonstrating precise classification reasoning. At the video-level, MeDiVLAD outperforms both LSTM~\cite{hochreiter1997long} and  NetVLAD~\cite{arandjelovic2016netvlad} aggregation, achieving superior performance in the video-level scoring. 

\section{Data \& Problem Statement}\label{sec:dataset}
\begin{figure}[tb]
  \centering
  \includegraphics[width=0.47\textwidth]{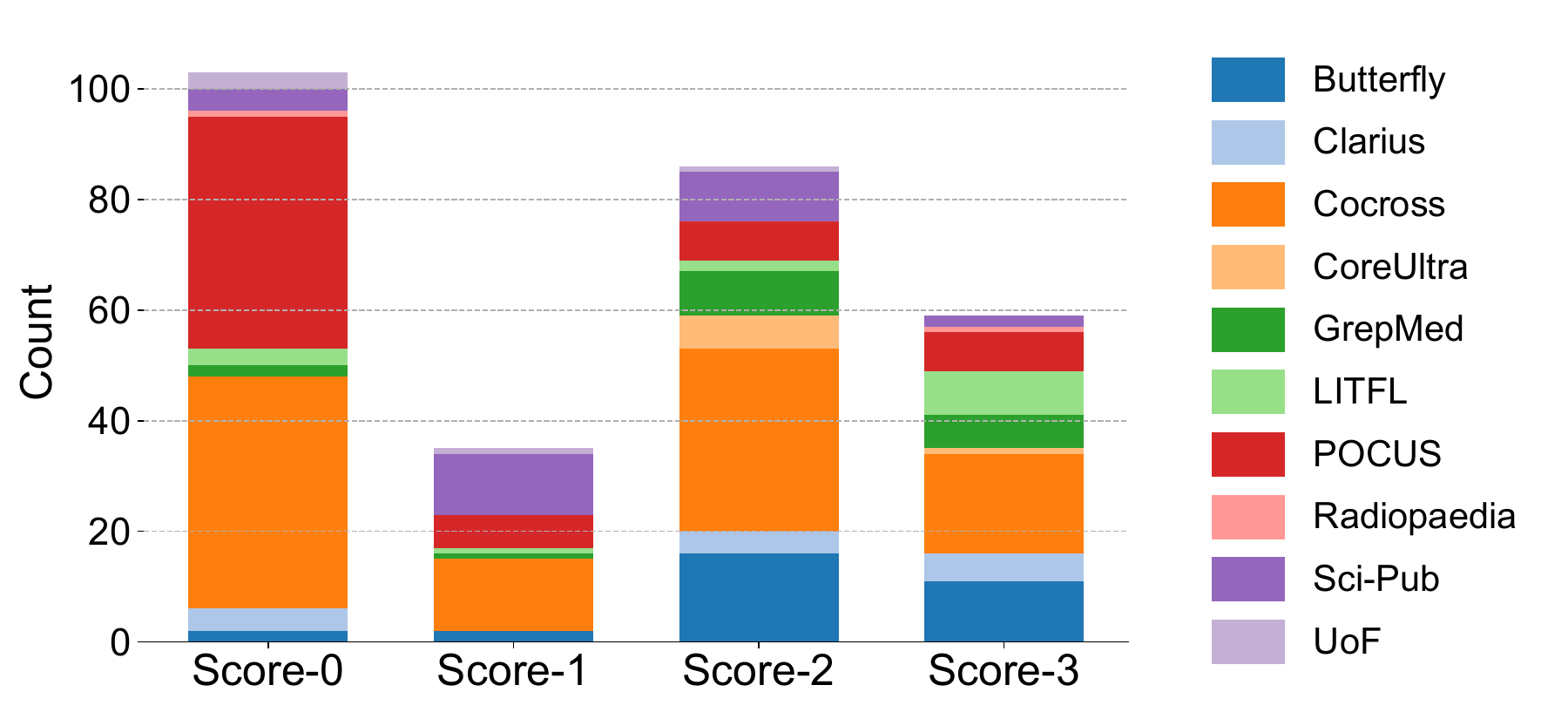}
  \hspace{-7pt}\vspace{-2pt}\caption{Score distribution of our LUS dataset, with different colors indicating different data sources. CoCross is our private dataset, other sources are detailed in~\cite{COVIDxUS2021}}
  \label{fig: data_distribution}
  \vspace{-5pt}
\end{figure}

We curated our LUS dataset with $177$ curvilinear ultrasound videos from the COVIDx-US~\cite{COVIDxUS2021} dataset and $106$ ultrasound videos collected from "G. Papanikolaou” General Hospital of Thessaloniki (CoCross), totaling $283$ videos from $156$ patients. The data distribution is shown in Fig~\ref{fig: data_distribution}. To improve frame-level scoring, we randomly selected $2-3$ representative frames from each video, resulting in a small frame-level dataset of $585$ annotated images.

We adopt an \textbf{improved LUS scoring system}, namely the integrated lung ultrasound score (i-LUS)\cite{dell2022integrated}. This system incorporates additional factors such as pleural line characteristics and cardiac involvement in COVID-19, providing a more comprehensive assessment. I-LUS uses a 4-level scoring system: \textbf{Score-0} represents a normal lung with a continuous pleural line and horizontal A-line artifact; \textbf{Score-1} indicates at least 2 isolated or coalescent B-lines covering less than 50\% of the image without clear sub-pleural alterations; \textbf{Score-2} includes B-lines covering more than 50\% of the image, still without clear sub-pleural alterations; and \textbf{Score-3} represents consolidation with poorly dynamic arborescent air bronchograms. Considering the extreme class imbalance in our dataset, \textbf{we combined scores 1 and 2, simplifying it to a 3-level scoring system.} Given such a three score system $\mathbf{Y} \in \{\mathbf{y}_0, \mathbf{y}_1, \mathbf{y}_2\}$ and a ultrasound video containing $\mathbf{N}$ frames, $\mathbf{V} \in \{\mathbf{x}_0, \allowbreak\mathbf{x}_1, \dots, \allowbreak \mathbf{x}_{N-1}\}$, our goal is to predict the probability $p(\mathbf{y}_i|\mathbf{x})$ for a single frame $\mathbf{x}$, and $p(\mathbf{y}_i|\mathbf{V})$ for the entire video $\mathbf{V}$. These probabilities are parameterized using two different neural networks $\phi_i$ and $\phi_v$, where $ \phi_i(\mathbf{x}) = p(\mathbf{y}_i|\mathbf{x})$ and $ \phi_v(\mathbf{x}_0, \mathbf{x}_1, \dots, \mathbf{x}_{N-1})=p(\mathbf{y}_i|\mathbf{V})$. For simplicity, all videos are reshaped to $224 \times 224$ and uniformly downsampled to $\mathbf{N} = 15$ frames. For samples with fewer than 15 frames, nearest-neighbor interpolation is applied to match the target frame count.

\section{Method}\label{sec:method}

\begin{figure*}[htb]
  \centering
  \includegraphics[width=0.98\textwidth]{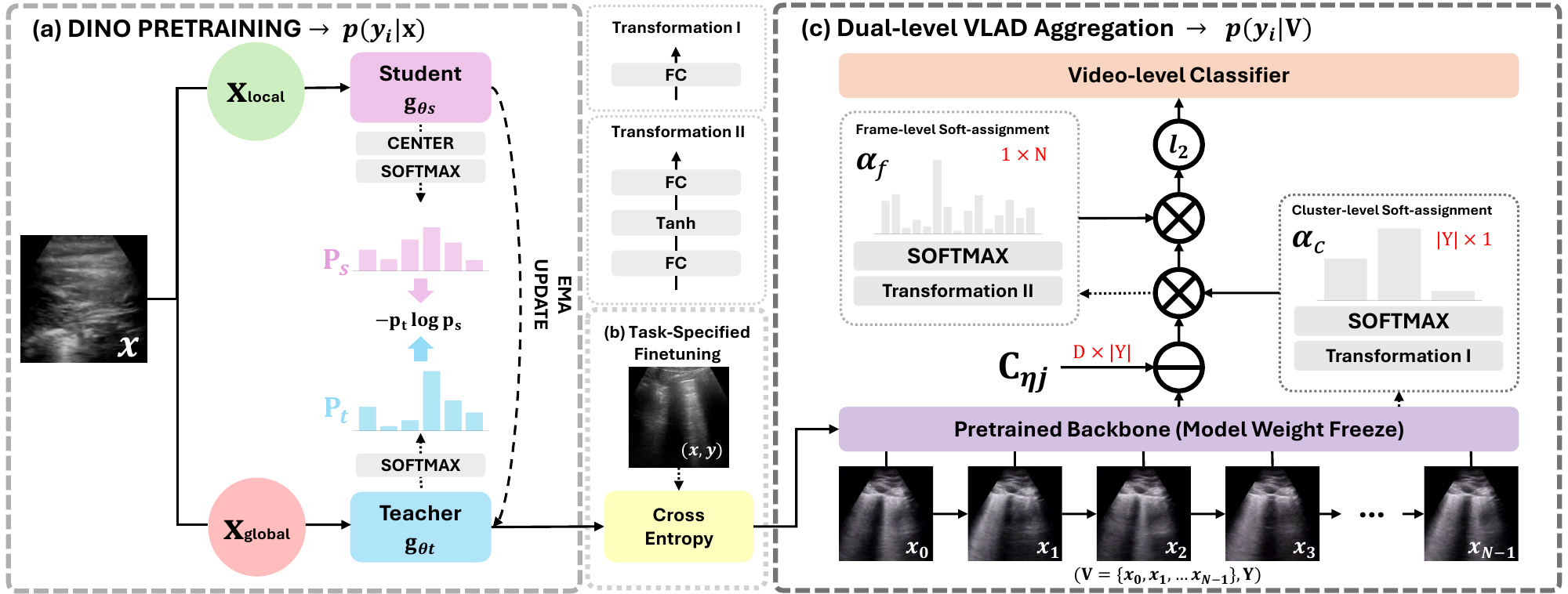}
  \hspace{-7pt}\vspace{-2pt}\caption{Overall structure of MeDiVLAD, including (a) DINO-based pretraining on unlabeled frames, (b) optional task-specific finetuning on a small set of labeled frames, and (c) dual-level VLAD aggregation for video-level severity scoring.}
  \vspace{-8pt}
  \label{fig: overall_structure}
\end{figure*}
\newcommand{\manualA}{\hyperref[fig: overall_structure]{2a}}
\newcommand{\manualB}{\hyperref[fig: overall_structure]{2b}}
\newcommand{\manualC}{\hyperref[fig: overall_structure]{2c}}
\subsection{Self-distillation with Task-specific Finetuning}
Transformers~\cite{vaswani2017attention} have recently emerged as an alternative to CNNs, offering superior performance in medical classification tasks~\cite{dai2021transmed, wu2023ctranscnn, manzari2023medvit}, with their inherent attention mechanisms providing precise reasoning. However, their limitations are also significant: they require more computational resources and training data, which restricts their applicability in most medical imaging scenarios. We question whether this issue can be mitigated by pretraining the neural network using the visual information contained in unlabeled images.

DINO~\cite{caron2021emerging} adopts a similar architecture to most recent SSL-based methods, with the key difference being its use of self-knowledge distillation during training. As shown in Fig~\manualA, DINO~\cite{caron2021emerging} leverages a teacher network $g_{\theta_t}$ to guide a student network $g_{\theta_s}$ with the same architecture, and $\theta_t$ and $\theta_s$ are both learnable parameters. During training, the input image $\mathbf{x}$ will be encoded into two sets of $\mathbf{K}$-dimensional distributions, $P_s$ and $P_t$. The key difference lies in that the teacher's input, $\{\mathbf{x}^g_i\}_{i=0}^m$, are more global than the student's input, $\{\mathbf{x}^l_j\}_{j=0}^n$. In practice, $\mathbf{x}^g$ refers to a larger crop of the original image $\mathbf{x}$, while $\mathbf{x}^l$ is a smaller, augmented version and we normally set $m>n$. The training objective~\cite{caron2021emerging} is to minimize the cross-entropy loss between every $P_s$ and $P_t$, thereby fostering the correspondence from local-to-global, i.e.,
\begin{equation}
\min _{\theta_s} \sum_{\mathbf{x} \in\left\{\mathbf{x}^g_i\right\}} \sum_{\mathbf{x} \in \left\{\mathbf{x}^l_j\right\}} H\left(P_t(\mathbf{x}^g, \tau_t), P_s\left(\mathbf{x}^l, \tau_s\right)\right)
\end{equation}
where $H(\star)$ is the cross-entropy loss and $P(\star)$ denotes the softmax operation with certain temperature $\tau$ which is a non-negative constant. During this process, only the student’s weights $g_{\theta s}$ are updated, while the teacher is updated via exponential moving average (EMA)~\cite{he2020momentum}, i.e., $\theta_t \leftarrow \lambda \theta_t+(1-\lambda) \theta_s$. Notably,~\cite{caron2021emerging} shows that \textbf{the teacher network produces better features} than the student. Therefore, we will use the teacher for subsequent tasks. It is worth mentioning that the encoded distributions $P_s$ and $P_t$ can be interpreted as the probabilities for the classification over $\mathbf{K}$ predefined classes, whereas our target classes are their subsets. To narrow down from the predefined classes toward LUS scoring, we further perform a fully supervised \textbf{task-specific finetuning} on the teacher network using a small set of annotated frames, which we illustrated in Fig.~\manualB. We denote the finetuned model as $g_{\theta_t}^*$.

\subsection{Dual-level VLAD Aggregation}
Under our default setting, each ultrasound video contains $\mathbf{N}=15$ frames, where each frame corresponding to an $\mathbf{D}$-dimensional feature vector $f$, extracted by $g_{\theta_t}^*$. For our proposed dual-level VLDA aggregation (Fig.~\manualC), we consider the simplest scenario by setting the number of clusters to be equal to the number of labels. Similar to the NetVLAD~\cite{arandjelovic2016netvlad} aggregation, we encode the frame-level ultrasound embedding into a $\mathbf{|Y|} \times \mathbf{D}$ dimensional feature vector $f_{i}^\prime$. This is done by assigning a learned cluster centroid $\mathbf{C}_{\eta_j}$ to every frame embedding and concatenating their residual:
\begin{equation}
f_{i}^\prime=\mathop{\|}\limits_{j=1}^{|\mathbf{Y}|}\left\{\alpha_c\left(f_i, j\right)\left(f_{i}-c_{j}\right)\right\}
\end{equation}
where $i \in\{1, \ldots, \mathbf{N}\}$, and $\alpha_c(\star)$ is a learnable cluster-level soft-assignment that assigns frames to clusters based on their proximity. NetVLAD~\cite{arandjelovic2016netvlad} directly measures the sum of $f_i$ across the frame level. However, LUS videos are typically scored based on the highest severity observed in the video. In other words, video-level scoring relies on a single or a few representative frames within the video. To address this, we introduce an additional frame-level soft-assignment denoted as $\alpha_f\left(f_i^\prime, \tau^\prime\right)$ to perform frame selection. This assignment is implemented through a simple multilayer perceptron (MLP) with tanh activation, where temperature $\tau^\prime$ was adopted to control the model’s focus on the most informative frame segments. Then, the video-level embedding $v$ is obtained by summing up weighted frame-level features and applying an intra-normalization $\sigma$ to suppress bursts~\cite{arandjelovic2013all}, i.e.,
\begin{equation}
    v = \sigma \left( \sum_i^N \alpha_f\left(f_i^\prime, \tau^\prime\right) f_i^\prime \right)
\end{equation}
Finally, a cross-entropy-based video-level classifier is applied to finish the video-level scoring.

\section{Experiments}
\begin{table}[htb]
\vspace{-15pt}
\caption{Experiment Implementation Details}\vspace{1pt}
\label{tab: implementation}
\centering
\resizebox{0.49\textwidth}{28pt}{
\setlength{\tabcolsep}{2.0pt}
\begin{tabular}{lccccc}
\toprule
Stage \rule{0pt}{1.8ex} & LR & WD & Epochs & Batch Size & Data\\ 
\midrule
Pretraining & 1.25e-4  & 0.1-0.5 & 30 & 64 & Unlabeled Frames \\
Finetuning & 5.00e-5 & 0.001 & 100 & 64 & Labeled Frames \\
VideoCls & 1.00e-3  & 0.00001 & 200 & 32 & Labeled Videos\\
\bottomrule
\end{tabular}}
\vspace{-8pt}
\end{table}
Given the objective of this work, we did not require a large amount of labeled frame data for training. As such, we performed a 2-fold validation, assigning 302 images from 136 videos to fold 1 and 283 images from 140 videos to fold 2, ensuring that videos from the same source did not appear in different folds. All experiments were conducted on a single Nvidia Quadro RTX 8000 GPU and optimized using the AdamW optimizer, with learning rates decaying according to a cosine schedule. For pretraining, we adopted the ViT-S/8 configuration and data augmentation setup from~\cite{caron2021emerging}, training the backbone by randomly sampling unlabeled frames from the ultrasound videos. The temperatures $\tau_t$ and $\tau_s$ were set to 0.5 and 0.1, respectively. Afterward, we finetuned the network on labeled frames in a fully supervised manner. At the video level, we trained the dual-level VLAD aggregation using labeled ultrasound videos. For simplicity, additional training details are summarized in Table~\ref{tab: implementation}.

\subsection{Image-level Classification}
\begin{table}[ht]
\vspace{-5pt}
\caption{Image-level Classification}\label{tab:image-level}
\centering
\resizebox{0.48\textwidth}{!}{
\setlength{\tabcolsep}{2.5pt}
\begin{tabular}{lcccccc}
\toprule
\textbf{Methods} \rule{0pt}{2.0ex} & \textbf{Backbone} & \textbf{Pretrain} & \textbf{ROC-AUC} & \textbf{k-NN} & \textbf{Linear} \\ 
\midrule
Sup. & ResNet50 & IMG & 0.793 & 55.77 & 63.92 \\
DINO~\cite{caron2021emerging} & ViT-S/8 & IMG & 0.866 & 57.85 & 70.57 \\
DINO~\cite{caron2021emerging} & ViT-S/8 & IMG/LUS & 0.878 & 63.36 & 75.05 \\
\bottomrule
\multicolumn{6}{l}{\textit{Supervised Finetuning on LUS Image Dataset}} \rule{0pt}{2.0ex}\\
\midrule
Sup. & ResNet50 & IMG & 0.863 & - & 71.94 \\
Scratch & ResNet50 & - & 0.786 & - & 64.08 \\
DINO~\cite{caron2021emerging} & ViT-S/8 & IMG & 0.900 & - & 78.30 \\
DINO~\cite{caron2021emerging} & ViT-S/8 & IMG/LUS & 0.917 & - & 82.47 \\
Scratch & ViT-S/8 & - & 0.702 & - & 58.85 \\
\bottomrule
\end{tabular}
}
\\
\raggedright 
{\vspace{2pt}\hspace{3pt} 
\scriptsize IMG: ImageNet dataset; \quad LUS: Lung-ultrasound dataset}\vspace{-8pt}
\end{table}

\begin{figure}[htb]
  \centering
  \includegraphics[width=0.48\textwidth]{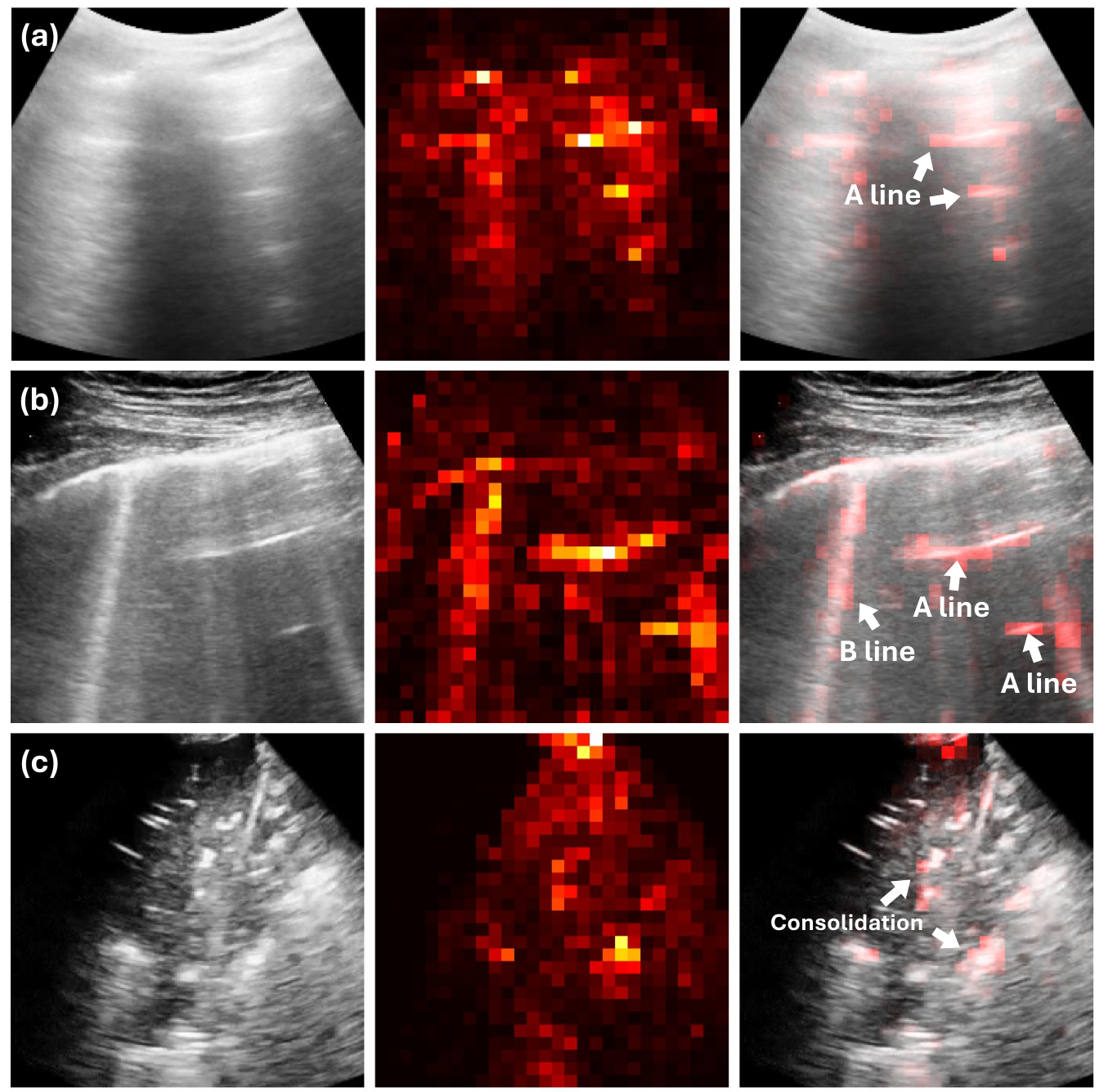}
  \vspace{-10pt}
\caption{Typical visualization of attention maps extracted from our pretrained ViT-S/8 (after finetuning).}
\vspace{-8pt}
\label{fig: attention map}
\end{figure}

We evaluated MeDiVLAD at the frame level. For this, we trained a ResNet-50 (23.5M) as a classification baseline with a similar number of parameters to the ViT-S (21.7M) we used. The average scoring accuracy (k-NN/linear classifier) and ROC-AUC (one-vs-all) were reported. It should be noted that the k-NN accuracy is only provided for the models that were not fully supervised during training. In the upper half of Table~\ref{tab:image-level}, we first examined the impact of self-distillation. Without using LUS data, ResNet-50 pretrained on ImageNet (IMG) showed slightly lower classification accuracy than the other two DINO experiment sets, with the advantage of DINO becoming more pronounced after incorporating unlabeled ultrasound data. As expected, after finetuning the model with labeled frames, both accuracy and ROC-AUC improved, outperforming all other baselines (AUC: $0.917$ \& Acc: $82.47\%$). Remarkably, even without supervised finetuning, we achieved an accuracy of $75.05\%$, surpassing the $71.94\%$ accuracy of the fully supervised ResNet-50. In Fig.~\ref{fig: attention map}, we present several attention map visualizations from the finetuned backbone. In (a) and (b), the attention maps accurately highlight both A-lines and B-lines, while in (c), the model identifies all regions of consolidation, offering clear insights into its decision-making process for LUS scoring.

\subsection{Video-level Classification}
\begin{table}[htb]
\vspace{-12pt}
\caption{Video-level Classification}\label{tab:video-level}\vspace{1pt}
\centering
\resizebox{0.48\textwidth}{!}{
\setlength{\tabcolsep}{2.5pt}
\begin{tabular}{lccccc}
\toprule
\textbf{Methods} \rule{0pt}{2.0ex} & \textbf{Backbone} & \textbf{Pretrain} & \textbf{ROC-AUC} & \textbf{Linear} \\ 
\midrule
GlobMAX~\cite{gare2022weakly} & ViT-S/8 & IMG/LUS$^*$ & 0.891 & 75.36 \\
Bi-LSTM~\cite{hochreiter1997long} & ViT-S/8 & IMG/LUS$^*$ & 0.897 & 78.59 \\
NetVLAD~\cite{arandjelovic2016netvlad} & ViT-S/8 & IMG/LUS$^*$ & 0.907 & 77.15 \\
\midrule
MeDiVLAD & ResNet50 & IMG/LUS$^*$ & 0.866 & 75.31 \\
MeDiVLAD & ViT-S/8 & IMG/LUS & 0.888 & 76.41 \\
MeDiVLAD & ViT-S/8 & IMG/LUS$^*$ & 0.936 & 82.60 \\
\bottomrule
\end{tabular}
}
\\
\raggedright 
{\vspace{2pt}\hspace{3pt} 
\scriptsize LUS$^*$: Finetuned in a supervised manner with the LUS image dataset.}
\end{table}

At the video level, we evaluated our proposed MeDiVLAD aggregation against three typical aggregation methods: Bi-LSTM~\cite{hochreiter1997long}, NetVLAD~\cite{arandjelovic2016netvlad}, and directly taking the max severity-category score~\cite{gare2022weakly} from the frame predictions. We used the same metrics as in the frame-level experiments for evaluation. For the LSTM, we simply set the hidden size to be the same as the embedding length. Additionally, we performed a grid search to find the best hyperparameters for each model and reported the metrics as the fold average. As shown in Tab.~\ref{tab:video-level}, MeDiVLAD significantly outperformed all other methods across all metrics. Similar to the results in the frame-level experiment, using our proposed dual-level VLAD aggregation, we achieved comparable performance to a finetuned ResNet-50, with an AUC of $0.86$ versus $0.88$ and an accuracy of $75.3\%$ versus $76.4\%$. These results not only confirm that applying self-knowledge distillation for pretraining a dedicated feature extractor is effective but also demonstrate that using a small amount of annotated data (300 samples) for finetuning can lead to an almost $6\%$ improvement in accuracy.

\section{Conclusion \& Discussion}
In this work, we introduced MeDiVLAD, a novel pipeline for LUS scoring at both frame and video levels. By leveraging self-knowledge distillation to pretrain a vision transformer without labels and using dual-level VLAD aggregation, we significantly reduced the reliance on expert annotation. At the frame level, our method achieved $75.05\%$ accuracy without labeled data, which improved to $82.47\%$ with finetuning. At the video level, MeDiVLAD outperformed other aggregation methods, such as Bi-LSTM and NetVLAD. These results highlight the outstanding performance of MeDiVLAD in LUS severity scoring, enabling it to support key applications such as the automatic identification of critical areas in severe lung pathology for further analysis. Furthermore, our pipeline offers a potential solution for broader medical imaging tasks, combining accuracy with interpretability in low-data settings.
\bibliographystyle{IEEEbib}
\bibliography{refs}

\end{document}